\documentclass[journal]{IEEEtran}
\usepackage{cite}

\ifCLASSINFOpdf
   \usepackage[pdftex]{graphicx}
\else
\fi
\usepackage{amsmath}
\usepackage{amssymb}

\DeclareMathOperator*{\argmin}{arg\,min}
\usepackage{bm}
\usepackage{makecell}
\usepackage{mathtools}
\usepackage{booktabs}
\usepackage{multirow}
\usepackage{floatrow}
\usepackage{subfigure}
\usepackage{wrapfig}
\usepackage[ruled]{algorithm2e}

\begin{document}
\begin{figure*}[t]
    \centering
    \includegraphics[width=\linewidth]{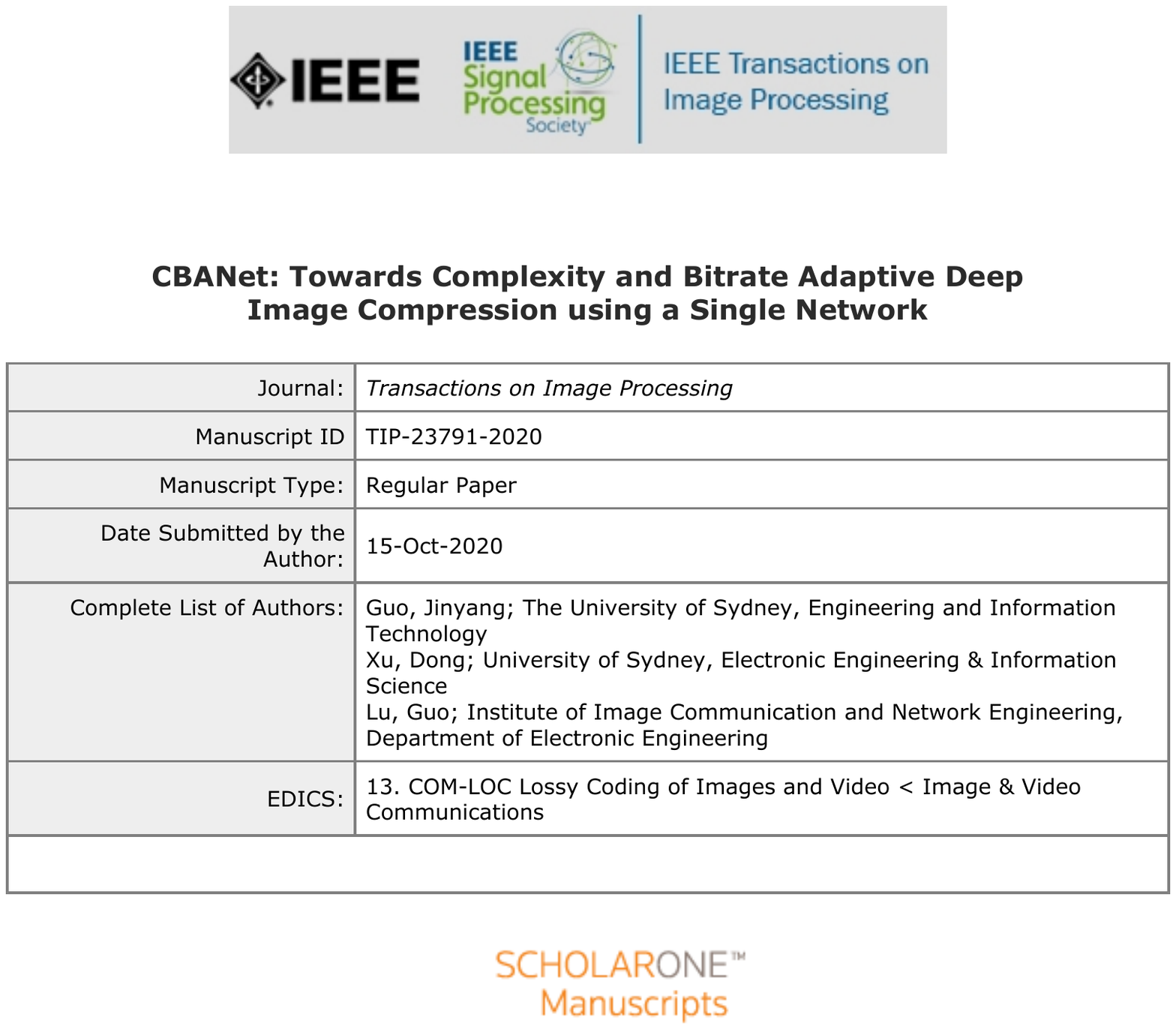}
\end{figure*}
%
\title{CBANet: Towards Complexity and Bitrate Adaptive Deep Image Compression using a Single Network}
%
%
%

\author{Jinyang~Guo,
        Dong~Xu,~\IEEEmembership{Fellow,~IEEE,}
        and~Guo~Lu
\thanks{Jinyang Guo and Dong Xu are with the School of Electrical and Information Engineering, The University of Sydney, NSW, Australia. E-mail: jinyang.guo@sydney.edu.au, dong.xu@sydney.edu.au}
\thanks{Guo Lu is with the School of Computer Science, Beijing Institute of Technology, Beijing, China. E-mail:  guo.lu@bit.edu.cn}
\thanks{Manuscript received April 19, 2005; revised August 26, 2015.}}

%
%

\markboth{Journal of \LaTeX\ Class Files,~Vol.~14, No.~8, August~2015}%
{Shell \MakeLowercase{\textit{et al.}}: Bare Demo of IEEEtran.cls for IEEE Journals}
%



\maketitle

\begin{abstract}
In this paper, we propose a new deep image compression framework called Complexity and Bitrate Adaptive Network (CBANet), which aims to learn one single network to support variable bitrate coding under different computational complexity constraints. In contrast to the existing state-of-the-art learning based image compression frameworks that only consider the rate-distortion trade-off without introducing any constraint related to the computational complexity, our CBANet considers the trade-off between the rate and distortion under \emph{dynamic} computational complexity constraints. Specifically, to decode the images with one single decoder under various computational complexity constraints, we propose a new multi-branch complexity adaptive module, in which each branch only takes a small portion of the computational budget of the decoder. The reconstructed images with different visual qualities can be readily generated by using different numbers of branches. Furthermore, to achieve variable bitrate decoding with one single decoder, we propose a bitrate adaptive module to project the representation from a base bitrate to the expected representation at a target bitrate for transmission. Then it will project the transmitted representation at the target bitrate back to that at the base bitrate for the decoding process. The proposed bit adaptive module can significantly reduce the storage requirement for deployment platforms. As a result, our CBANet enables one single codec to support multiple bitrate decoding under various computational complexity constraints. Comprehensive experiments on two benchmark datasets demonstrate the effectiveness of our CBANet for deep image compression.
\end{abstract}


%
\IEEEpeerreviewmaketitle

\section{Introduction}
\IEEEPARstart{T}{he} image compression techniques are widely used to decrease the number of bits required for storage and transmission. In recent years, the learning based image compression approaches~\cite{balle2017end,balle2018variational,minnen2018joint} achieved state-of-the-art performance by using the rate-distortion optimization technique. Although these methods have achieved significant success in this field, it is still a challenging task to deploy these learning based image compression methods in real-world applications. 

\begin{figure*}[t]
    \centering
    \includegraphics[width=\linewidth]{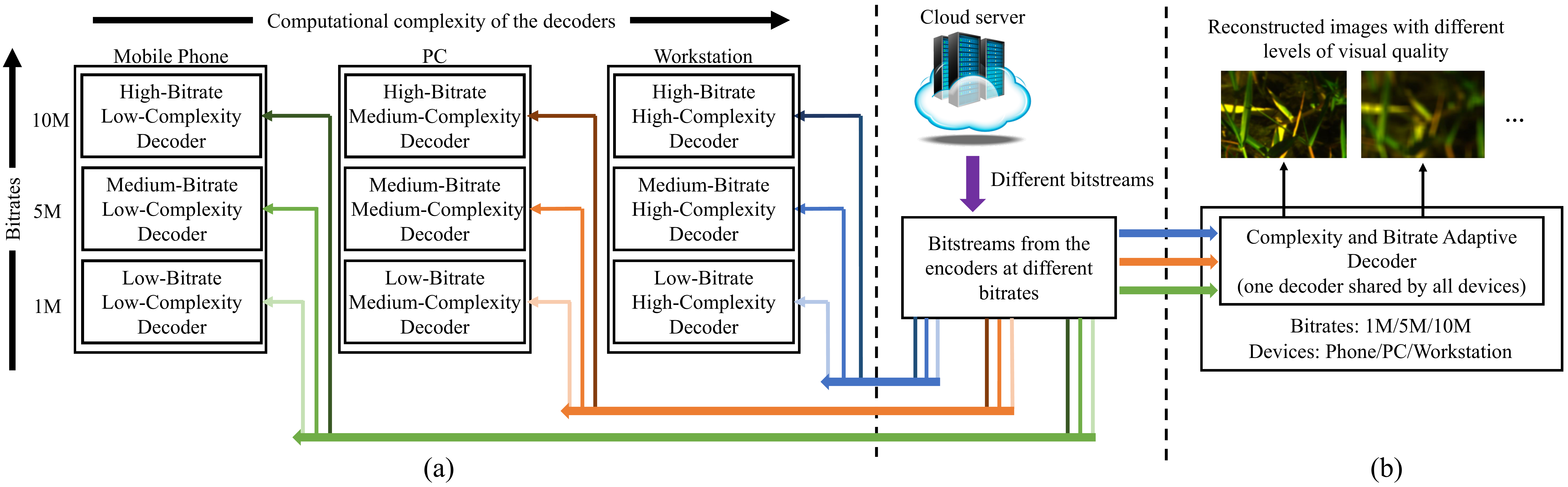}
    \caption{Comparison of (a) the existing learning based image compression framework, which requires multiple decoders for decoding the images at different bitrates on different devices with different computational complexity levels and (b) our CBANet, which can decode the bitstreams at different bitrates on different devices with different computational complexity constraints by using one single decoder.}
    \label{fig:intro}
\end{figure*}

First, the computational complexity constraints for the learning based image compression systems often change when deploying the image codecs on different platforms. These platforms can range from a cluster of servers with sufficient computational resources to a tiny chip with limited computational resources. Besides, the allocated computational resources for one deployment platform may also change over time (e.g., when the battery status of mobile phones changes). However, the network architecture of the existing learning based image compression frameworks like \cite{balle2017end, balle2018variational, minnen2018joint} is fixed at the inference stage. In other words, we have to design and store multiple codecs for different deployment scenarios in order to satisfy the requirements of any specific deployment platforms. Therefore, it is useful to develop an image compression system that can support different computational complexity levels by using a single model for practical applications.

Second, a learning based image compression framework that can support variable bitrate coding is also required when the bandwidth condition changes over time. However, the current state-of-the-art learning based image compression frameworks \cite{balle2017end, balle2018variational, minnen2018joint} require different models for different bitrates, which increases the storage requirement for these methods. Although several recent methods have been proposed~\cite{choi2019variable,lu2020end} to support variable bitrate coding, these works do not consider dynamic computational budgets. Therefore, it is also desirable to develop a compression system that can support various bitrates under dynamic computational complexity constraints by using a single model.

As shown in Fig.~\ref{fig:intro}(a), we are often given dynamic computational complexity constraints and changing bandwidth conditions in real-world deployment scenarios. A simple approach to support these deployment scenarios is to train multiple learning based codecs at different bitrates and under various computational complexity constraints. However, it is obvious that the required number of decoders is equal to the number of computational complexity constraints multiply the number of bitrates, which significantly increases the storage cost for saving multiple models. 

Therefore, a new research problem emerges: \textit{How to design an image compression system to support multiple bitrates under dynamic computational complexity constraints by using a single network?} 
In this work, we propose a new learning based image compression framework called Complexity and Bitrate Adaptive Network (CBANet), which aims to use one single model to support variable bitrate coding under different computational complexity constraints. Considering that the decoder side is more sensitive to the computational complexity or the storage requirement, we only focus on the decoder side in order to address this problem in a more efficient way. After finishing the training process, our CBANet can reconstruct the images at different bitrates and different computational complexity constraints without performing additional re-training process (see Fig.~\ref{fig:intro}(b)). Specifically, our CBANet consists of two modules: a complexity adaptive module~(CAM) and a bitrate adaptive module~(BAM). The CAM consists of several parallel branches, in which each branch takes a small portion of the computational budget and the reconstructed images with different visual qualities can be readily generated by using different numbers of branches with different computational costs. To support variable bitrate coding with one single model, the BAM uses a small convolutional neural network~(CNN) to project the representation at a base bitrate to the expected representation at a target bitrate for transmission. Then it will convert the representation at the target bitrate to that at the base bitrate for the decoding process of the CAM. By jointly considering the rate-distortion-complexity trade-off in our newly proposed optimization problem and seamlessly integrating the proposed CAM and BAM into a single network, we can use one deep model to support multiple bitrates under various computational complexity constraints, which helps us to deploy the learning based image codecs in different scenarios for practical applications.

To the best of our knowledge, this is the first deep image compression system that can learn a single deep model to support various bitrates under dynamic computational complexity constraints. Comprehensive experiments on two benchmark datasets demonstrate the effectiveness of our CBANet for image compression.

\section{Related Work}
\subsection{Image Compression}
In the past decades, a large number of conventional image compression standards~\cite{wallace1992jpeg,skodras2001jpeg,bellard2018bpg} were proposed. However, these compression standards heavily rely on the hand-crafted techniques. In recent years, several learning based image compression  methods~\cite{toderici2015variable,balle2017end,balle2018variational,toderici2017full,agustsson2017soft,theis2017lossy,mentzer2018conditional,minnen2018joint,agustsson2019generative,patel2019deep,lee2018context,blau2019rethinking,li2018learning,rippel2017real} were proposed to improve the compression performance, among which the convolutional neural network (CNN) based methods are attracting increasing attention. For example, Ball{\'e} et al.~\cite{balle2017end} proposed an end-to-end optimized image compression system by minimizing the rate-distortion trade-off~\cite{shannon2001mathematical}. It was subsequently extended in \cite{balle2018variational} by additionally introducing the hyperprior model. To achieve better bitrate estimation, the method in \cite{minnen2018joint} further extended the work in \cite{balle2018variational} by introducing the context model for the adaptive arithmetic coding.

\subsection{Variable Bitrate Image Compression}
Several approaches~\cite{toderici2015variable,toderici2017full,johnston2018improved} were recently proposed to achieve variable bitrate compression. For example, Toderici et al.~\cite{toderici2015variable,toderici2017full} proposed the recurrent neural network (RNN) based image compression systems, where the reconstructed images at different bitrates can be generated at different iterations in RNN (i.e., with different computational complexity levels). However, the RNN based methods usually ignore the bitrate term in the optimization procedure and cannot support variable bitrate coding at any given computational complexity constraints. In other words, the computational complexity level increases as the bitrate increases. As a result, at one bitrate, these methods can only decode the bitstream under one fixed computational complexity level instead of supporting multiple computational complexity levels as our CBANet. On the other hand, several CNN based methods were also proposed~\cite{choi2019variable,lu2020end,yang2020variable,akbari2020learned} to support variable bitrate coding for the learning based image and video compression methods. However, these works do not consider dynamic computational budgets. When different computational budgets are provided, these methods need to train multiple deep models, which significantly increases the storage cost. In contrast, our work can support variable bitrate coding under dynamic computational complexity constraints by using one single model, which cannot be achieved by the existing works.

\begin{table}[t]
    \small
	\centering
	\caption{Difference between our CBANet and other image compression methods. RD represents the trade-off between the rate and distortion, while RDC represents the trade-off among the rate, distortion and computational complexity. MB and MC represent multiple bitrates and multiple computational complexity levels, respectively. * denotes the method cannot support multiple computational complexity levels at any given bitrate or support multiple bitrates at any given computational complexity level.}
    \begin{tabular}{c||cccc}
    \toprule[1pt]
     & RD & RDC & MB & MC \\
    \midrule[1pt]
    RNN methods~\cite{toderici2015variable,toderici2017full} & & & \checkmark* & \checkmark*\\
    CNN methods~\cite{balle2017end,balle2018variational,minnen2018joint} & \checkmark & & &\\
    Variable bitrate~\cite{choi2019variable,lu2020end} & \checkmark & & \checkmark &\\
    Johnston et al.~\cite{johnston2019computationally} &  & \checkmark & &\\
    Ours & & \checkmark & \checkmark & \checkmark \\
    \bottomrule[1pt]
    \end{tabular}
    \label{tab:relatedwork}
\end{table}

\subsection{Efficient Image Compression}
To reduce the computational complexity of the learning based image compression methods, Johnston et al.~\cite{johnston2019computationally} proposed to jointly consider the trade-off between the rate, distortion and the computational complexity, and select the optimal network architecture for each given computational budget. However, this approach has to train multiple models for multiple computational complexity levels or different bitrates. In contrast, our CBANet can well support \textit{dynamically} changed computational budget through one single model. 

To better understand our approach, we summarize the difference between our CBANet and the existing image compression methods in Table~\ref{tab:relatedwork}. From Table~\ref{tab:relatedwork}, our CBANet can support both multiple bitrates and multiple computational complexity levels, which cannot be achieved by the existing learning based deep image compression systems.

\subsection{Dynamic Neural Networks}
Many works~\cite{wu2018blockdrop,liu2018dynamic,wang2018skipnet,lin2017runtime,yu2019slimmable,yu2019universally,cai2019once} were proposed to dynamically construct the networks in order to deploy CNNs at different scenarios. For example, Yu et al.~\cite{yu2019slimmable} proposed a slimmable network structure to process the images by adaptively choosing the networks with different numbers of channels. Cai et al.~\cite{cai2019once} proposed the once-for-all network structure to deploy the network at different scenarios. However, these approaches only focus on the image classification task and do not investigate the more challenging image compression task.
In the image compression task, the input bitstreams will dynamically change if we aim to decode the bitstreams at various bitrates. If we simply apply these methods for the image compression task, we still require different CNNs at different bitrates. In contrast to these approaches, our CBANet can use a single CNN for decoding the bitstreams at multiple bitrates under various computational complexity constraints.

\section{Methodology}
\label{Sec:Methodology}
In this section, we use the method \cite{balle2017end} as an example to illustrate our CBANet, which is the most fundamental structure for the learning based image compression framework. We firstly introduce the overview of our CBANet. Then we will describe our complexity adaptive module (CAM) and bitrate adaptive module (BAM) in details. Finally, we will discuss the detailed structures of our CBANet based on other methods.

\subsection{Overview}
\textbf{Formulation.}
For the learning based image compression framework, the auto-encoder structure is the most popular one~\cite{balle2017end,balle2018variational,minnen2018joint}, which consists of an encoder and a decoder. Given the input image $x$, the encoder extracts the latent representation, which will then be quantized. The decoder takes the quantized representation as the input and uses a series of deconvolution operations to generate the reconstructed frame. The goal of the existing learning based image compression codec is to minimize the distortion $D$ for any given bitrate $R_{giv}$, i.e.,
\begin{equation}
    {\rm min} \ D,\ s.t. \ R \leq R_{giv},
    \label{eqn:rd}
\end{equation}
where $R$ is the coding bitrate. This rate-distortion trade-off is usually solved by using the Lagrangian multiplier optimization technology, in which we usually train multiple models by using different pre-defined Lagrangian multiplier values.

However, the rate-distortion trade-off in Eq.~(\ref{eqn:rd}) does not consider the computational complexity constraint. In contrast, we aim to minimize the following objective function in this work:
\begin{equation}
   {\rm min}\ D,\ s.t. \ R \leq R_{giv}, \ C \leq C_{giv},
    \label{eqn:rdc}
\end{equation}
where $C$ represents the computational complexity of the current codec and $C_{giv}$ is the given computational complexity constraint. 

\begin{figure}[t]
\centering
\includegraphics[width=\textwidth]{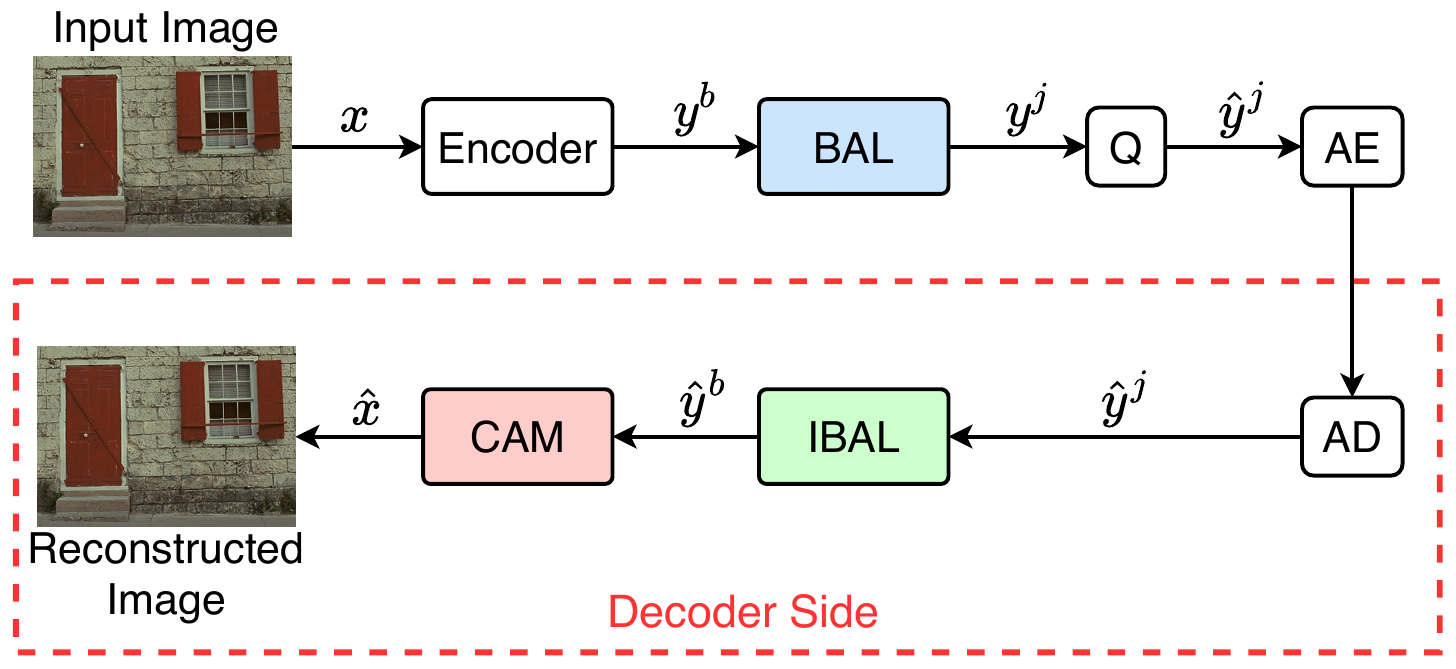}
\caption{Illustration of our CBANet, in which we take the method in \cite{balle2017end} as an example for illustration. ``AE'' and ``AD'' represent the arithmetic encoder and arithmetic decoder, respectively. ``Q'' represents the quantization operation. ``BAL'' and ``IBAL'' represent the bitrate adaptive layer and the inverse bitrate adaptive layer in our newly developed bitrate adaptive module (BAM), respectively. ``CAM'' represents the complexity adaptive module in our CBANet.}
\label{fig:overview}
\end{figure}

\textbf{Our overall pipeline.}
In this work, we focus on the decoder side, in which we aim to support multiple bitrates and computational complexity levels by using a single model. As shown in Fig.~\ref{fig:overview}, our proposed CBANet consists of two modules: a bitrate adaptive module (BAM) and a complexity adaptive module (CAM). Specifically, the BAM consists of two parts: the bitrate adaptive layer (BAL) and the inverse bitrate adaptive layer (IBAL). The BAL aims to transfer the representation $y^{b}$ at a base bitrate to that at the $j$-th supported bitrate $y^{j}$. After the quantization operation, the quantized representation $\hat{y}^{j}$ will be transmitted to the decoder side and we use the IBAL to transfer $\hat{y}^{j}$ back to the representation at the base bitrate $\hat{y}^{b}$. Then we feed $\hat{y}^{b}$ to the CAM, which can decode the images at different computational complexity levels. Finally, our CBANet can support multiple bitrate decoding under various computational complexity levels through a single model. We will introduce more details about the BAM and the CAM below.

\textbf{The optimization procedure.}
Let us assume the trainable parameters of our CBANet are represented as $\bm{\theta}$, which consists of the parameters from two modules: $\bm{\theta}^{BAM}$ from the BAM and $\bm{\theta}^{CAM}$ from the CAM. Note $\bm{\theta}^{BAM}$ consists of the parameters from both the BAL (i.e., $\bm{\theta}^{BAL}$) and the IBAL (i.e., $\bm{\theta}^{IBAL}$). It is a non-trivial task to solve the optimization problem in Eq.~(\ref{eqn:rdc}) by learning all the parameters $\bm{\theta}$ in one step. Instead, in our proposed framework, we disentangle the parameters of these two modules and propose a two-step optimization procedure to respectively learn $\bm{\theta}^{CAM}$ and $\bm{\theta}^{BAM}$ at each step. Specifically, at the first step, we propose to optimize the parameters $\bm{\theta}^{CAM}$ in CAM by solving the following objective function related to the distortion and complexity trade-off when the bitrate is fixed:
\begin{equation}
    \argmin_{\bm{\theta}^{CAM}} D,\ s.t. \ C = C_i, \ R= R_{b}, \ \forall \ i=1, \dots ,N,
    \label{eqn:D1}
\end{equation}
where $D$ denotes the distortion. $R$ and $C$ denote the bitrate and computational complexity level, respectively. $N$ is the total number of the supported computational complexity levels. $R_{b}$ and $C_i$ are the base bitrate and the $i$-th supported complexity level in our codec, respectively. In our implementation, we use the highest supported bitrate as our base bitrate $R_{b}$. In this optimization procedure, we aim to learn a complexity adaptive decoder, which can readily generate the reconstructed images with different visual qualities based on different computational complexity levels $C_{i}$ at the fixed bitrate $R_{b}$.

At the second step, we fix the CAM and train the BAM by solving the following objective function related to the distortion and bitrate trade-off:
\begin{equation}
    \argmin_{\bm{\theta}^{BAM}} D,\ s.t. \ C =  C_{b}, \ R \leq R_{j}, \ \forall \ j = 1, \dots, M-1,
    \label{eqn:D2.1}
\end{equation}
where $C_{b}$ is the base computational complexity level of our CAM and $R_{j}$ is the $j$-th supported bitrate. $M$ is the total number of the supported bitrates. We solve the optimization problem in Eq.~(\ref{eqn:D2.1}) by using the Lagrangian multiplier optimization technology. Therefore, the optimization problem in Eq.~(\ref{eqn:D2.1}) can be rewritten as follows:
\begin{equation}
\begin{aligned}
    &\argmin_{\bm{\theta}^{BAM}}  D + \lambda_j^{'} R, \text{ or equivalently, } \argmin_{\bm{\theta}^{BAM}}  R + \lambda_j D, \\
    &s.t. \ C =  C_{b}, \ \forall \ j = 1, \dots, M-1,
    \label{eqn:D2.2}
\end{aligned}
\end{equation}
where $\lambda_{j}^{'}=1/\lambda_{j}$ and $\lambda_{j}$ is the $j$-th Lagrangian multiplier to produce the BAM that can support the $j$-th bitrate. The other notations are the same as those in Eq.~(\ref{eqn:D2.1}). In our implementation, we use the highest supported computational complexity level as the base computational complexity level $C_{b}$. Note when $j=M$, we directly set $y^{j}=y^{b}$ and $\hat{y}^{b}=\hat{y}^{j}$ by using the identity mapping in both the BAL and the IBAL (see Sec.~\ref{Sec:BAM} for more details). At the second step, given the fixed computational complexity level, we train the BAM for different bitrates in order to support variable bitrate coding. After the training process is finished, our BAM can transfer the representation $y^{b}$ at the base bitrate to that at the $j$-th supported bitrate for transmission and transfer the transmitted representation $\hat{y}^{j}$ back to that at the base bitrate $\hat{y}^{b}$ for the decoding process of CAM.

\textbf{Discussion.} The two objective functions in Eq.~(\ref{eqn:D1}) and Eq.~(\ref{eqn:D2.1}) are proposed based on the following two observations: 
(1) When the bitrate $R$ is fixed as the base bitrate (i.e., $R=R_{b}$), there is no reconstruction loss from the BAM as the output and the input of the BAM are the same (see Sec.~\ref{Sec:BAM}). Therefore, we focus on achieving the distortion and complexity trade-off in Eq.~(\ref{eqn:D1}) without considering the influence from the BAM. 
(2) On the other hand, our CBANet at the highest computational complexity level achieves the best compression performance (see Sec.~\ref{Sec:Experiments} for more details). Therefore, we use the highest computational complexity level as our base computational complexity level $C_{b}$ and then train the BAM to focus on achieving the rate-distortion trade-off in Eq.~(\ref{eqn:D2.1}), in which the reconstruction loss from the CAM will not significantly affect the training process of the BAM. 
In summary, instead of directly solving Eq.~(\ref{eqn:rdc}) in this work, we achieve an alternative rate-distortion-complexity trade-off by using a two-step optimization strategy. Our training strategy disentangles the two sets of parameters and thus the training process for learning one set of parameters (i.e., $\bm{\theta}^{CAM}$ or $\bm{\theta}^{BAM}$) is relatively independent and will not be substantially affected by the reconstruction loss from another module.

\subsection{Encoder Preparation}
For the preparation of the encoder, we follow the existing learning based image compression methods like \cite{balle2017end,balle2018variational,minnen2018joint} to train one pair of encoder-decoder at the base bitrate. Then, we use the learned encoder to generate the bitstream at the base bitrate $y^{b}$. Note that once the encoder is trained at this stage, we will fix the encoder in the remained training procedure. Our CBANet based decoder will replace the original decoder in order to support multiple bitrate decoding under various computational complexity constraints by using a single model.

\subsection{Complexity Adaptive Module (CAM)}
\begin{figure}[t]
\centering
\includegraphics[width=\textwidth]{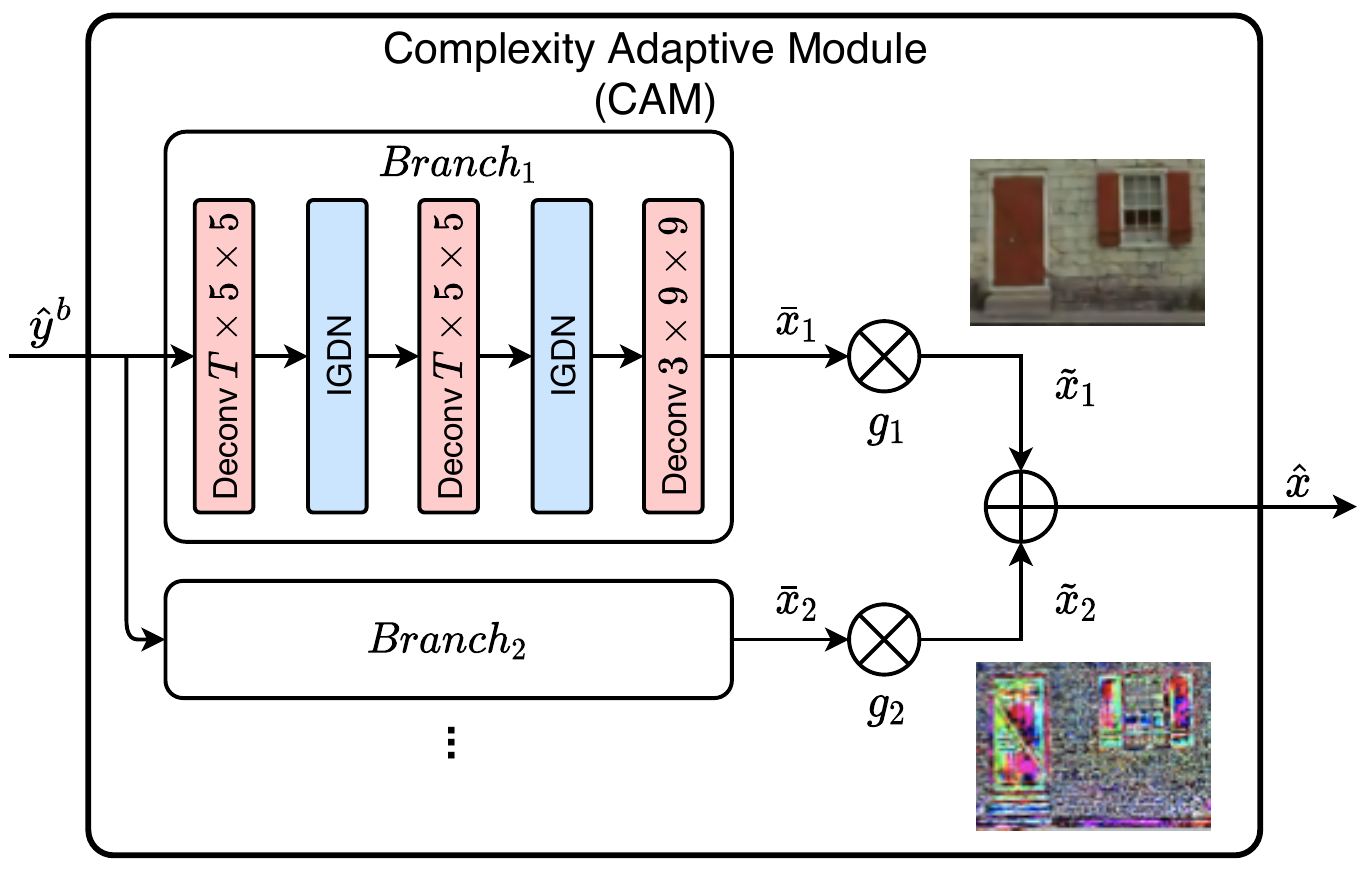}
\caption{Illustration of our complexity adaptive module (CAM). ``Deconv'' represents the deconvolutional layer. Its parameter is denoted as: \#output channels $\times$ filter height $\times$ filter width. ``IGDN'' represents the inverse generalized divisive normalization layer~\cite{balle2017end}. $T$ is the number of output channels at each layer. $\oplus$ is the element-wise summation and $\otimes$ is the element-wise multiplication.}
\label{fig:CAM}
\end{figure}

Fig.~\ref{fig:CAM} shows the structure of our CAM, which consists of several parallel branches. Each branch is a CNN that directly takes the representation $\hat{y}^{b}$ (it will be introduced in Eq.~(\ref{eqn:IBAL})) as the input and generates one residual component of the reconstructed images. Formally, the output of the $k$-th branch $\bar{x}_{k}$ can be formulated as follows:
\begin{equation}
    \bar{x}_{k} = \mathcal{F}_{k}(\hat{y}^{b}; \bm{\theta}^{CAM}_{k}),
    \label{eqn:SD}
\end{equation}
where $\mathcal{F}_{k}(\cdot, \cdot)$ is the function from the decoding process for the $k$-th branch in CAM and $\bm{\theta}^{CAM}_{k}$ denotes the parameters of the $k$-th branch of the CAM. After producing the output representation from all the branches, we aggregate the outputs from different branches by using the weighted sum of them, which can be written as follows:
\begin{equation}
    \hat{x} = \sum^{K}_{k=1}\tilde{x}_{k} = \sum^{K}_{k=1}g_{k} \cdot \bar{x}_{k},
    \label{eqn:aggregate}
\end{equation}
where the scalar $g_{k}$ is the weight for the $k$-th branch and it will be automatically learned in the training process. $\tilde{x}_{k}$ is the scaled output of the $k$-th branch and $K$ is the number of branches used in the decoding process. We have $K \leq K_{max}$ where $K_{max}=3$ in our implementation is the pre-defined total number of branches. In the decoding process, the computational complexity level will increase when the number of used branches $K$ increases. $\hat{x}$ is the final reconstructed image after aggregating the outputs from $K$ branches. Based on the specific computational complexity constraint of the device/platform, we can automatically choose the number of used branches in the decoding process to generate the reconstructed images with different visual qualities.

\textbf{Details of the training process.}
We train the CAM by solving the objective function in Eq.~(\ref{eqn:D1}) in a progressive manner. Specifically, we use the representation $\hat{y}^{b}$ at the base bitrate (i.e., $R=R_{b}$) as the input of the CAM and train the first branch by setting $K=1$ in Eq.~(\ref{eqn:aggregate}). We minimize the distortion between the reconstructed image and the input image and then back-propagate the gradients to update the parameters of the first branch. After that, we fix the parameters of the first branch and train the second branch (i.e., we set $K=2$ in Eq.~(\ref{eqn:aggregate})) in the same way as that for the first branch. We repeat this procedure in a branch-by-branch fashion and learn the parameters in all branches of our complexity adaptive module. Finally, we obtain $\bm{\theta}^{CAM}$, which consists of the learned parameters from all the branches.

\subsection{Bit Adaptive Module (BAM)}
\label{Sec:BAM}
\begin{figure}[t]
\centering
\includegraphics[width=0.8\textwidth]{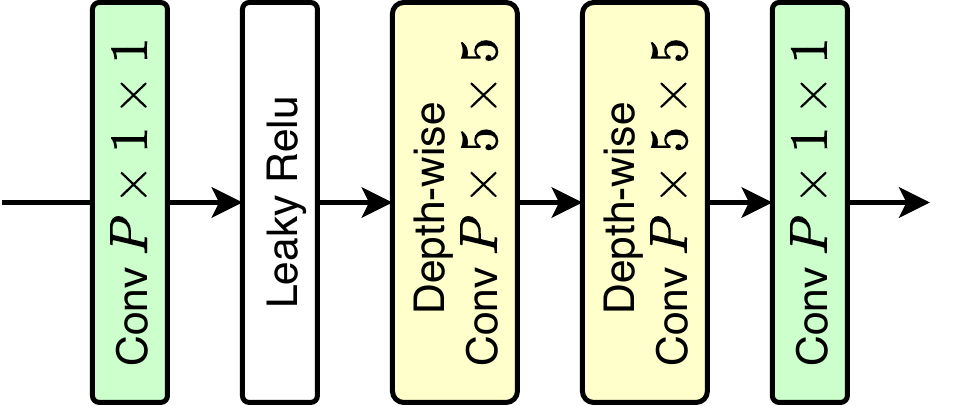}
\caption{Structure of the four-layer CNN in our bitrate adaptive layer (BAL) and inverse bitrate adaptive layer (IBAL). ``Conv'' and ``Depth-wise Conv'' represent the convolutional layer and the depth-wise convolutional layer, respectively. Their parameters are denoted as: \#output channels $\times$ filter height $\times$ filter width. ``Leaky Relu'' represents the leaky relu activation function. $P$ is the number of output channels at each layer.}
\label{fig:BAM}
\end{figure}

Our bitrate adaptive module (BAM) is used to achieve variable bitrate coding, which consists of two parts: the bitrate adaptive layer (BAL) and the inverse bitrate adaptive layer (IBAL). The BAL aims to transfer the representation from the base bitrate to that at the $j$-th supported bitrate for transmission, while the IBAL aims to transfer the transmitted representation from the $j$-th supported bitrate to that at the base bitrate for the decoding process of CAM. Inspired by \cite{lu2020end}, we use a four-layer CNN structure to implement our BAL and IBAL. As shown in Fig.~\ref{fig:overview}, given the input representation of the BAL $y^{b}$, the output of BAL $y^{j}$ can be written as follows:
\begin{equation}
    y^{j} = \left\{
    \begin{aligned}
    &y^{b} &\text{when } R=R_{b},\\
    &y^{b} \cdot  (1 - Sigmoid(\mathcal{B}(y^{b}; \bm{\theta}^{BAL}))) &\text{when } R \neq R_{b},
    \end{aligned}
    \right.
    \label{eqn:BAL}
\end{equation}
where $\mathcal{B}(\cdot, \cdot)$ is the function of our four-layer CNN for BAL and its parameters is denoted as $\bm{\theta}^{BAL}$. $R$ is the bitrate of the input representation and $R_{b}$ is the base bitrate. 

Similar to the BAL, given the input representation of the IBAL $\hat{y}^{j}$, the output representation of IBAL $\hat{y}^{b}$ can be written as follows:
\begin{equation}
    \hat{y}^{b} = \left\{
    \begin{aligned}
    &\hat{y}^{j} &\text{when } R=R_{b},\\
    &\hat{y}^{j} \cdot  (1 + Relu(\hat{\mathcal{B}}(\hat{y}^{j}; \bm{\theta}^{IBAL}))) &\text{when } R \neq R_{b},
    \end{aligned}
    \right.
    \label{eqn:IBAL}
\end{equation}
where $\hat{\mathcal{B}}(\cdot, \cdot)$ is the function of the four-layer CNN for IBAL and its parameters is denoted as $\bm{\theta}^{IBAL}$. The other notations are the same as those in Eq.~(\ref{eqn:BAL}). We use the same structure as in \cite{lu2020end} to implement the four-layer CNN for BAL and IBAL, which is illustrated in Fig.~\ref{fig:BAM}.

\textbf{Details of the training process.}
We train one BAM for each bitrate by solving the objective function in Eq.~(\ref{eqn:D2.2}). Specifically, we feed the output of our encoder $y^{b}$ to the subsequent BAL and IBAL to generate the representation $\hat{y}^{b}$ at the decoder side. Then we feed $\hat{y}^{b}$ to the CAM with all branches and obtain the final reconstructed image. We optimize the objective function in Eq.~(\ref{eqn:D2.2}) and back-propagate the gradients to update the parameters in BAM while keeping the parameters from the CAM and the encoder fixed. In this way, we can optimize the parameters in BAM in an end-to-end manner. 

Since the proposed decoder should support multiple bitrates and the quantized representation are quite different at different bitrates, we train one BAM for each bitrate except for the base bitrate (i.e., $R_{b}$ in Eq.~(\ref{eqn:D1})). Considering that the introduction of BAM only slightly increases the number of trainable parameters and the computational complexity (see Sec.~\ref{Sec:Experiments} for more details), our approach can support the multiple bitrate setting without significantly increasing the storage requirement when deploying our CBANet in various deployment scenarios.

\subsection{Structures of Our CBANet based on Other Methods}
\begin{figure}[t]
\centering
\includegraphics[width=\textwidth]{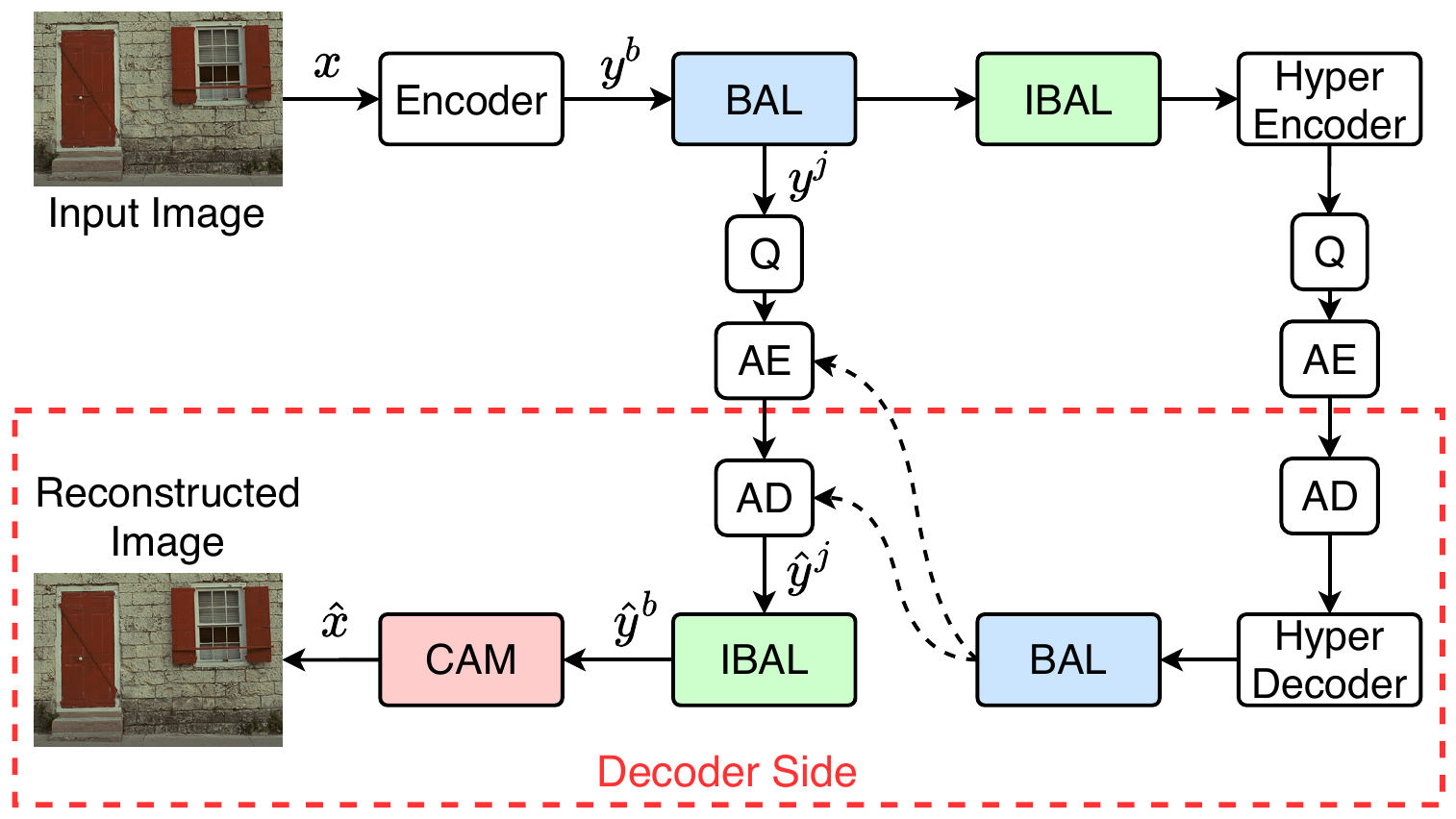}
\caption{Structure of our CBANet-HP, which is built upon the baseline method Ball{\'{e}} et al. (2018)~\cite{balle2018variational}. We additionally introduce one bitrate adaptive layer (BAL) and one inverse bitrate adaptive layer (IBAL) for the hyperprior part (i.e., ``Hyper Encoder'' and ``Hyper Decoder'').}
\label{fig:iclr18}
\end{figure}
\textbf{Structure of CBANet-HP: our CBANet based on Ball{\'{e}} et al. (2018).}
For better presentation, we name our CBANet based on the baseline method Ball{\'{e}} et al. (2018)~\cite{balle2018variational} as \textit{CBANet-HP}. Fig.~\ref{fig:iclr18} shows the network structure of our CBANet-HP. Similar to the encoder in our CBANet, the parameters of the hyperprior part (i.e., Hyper Encoder and Hyper Decoder in Fig.~\ref{fig:iclr18}) in our CBANet-HP are directly copied from the learned baseline model at the base bitrate, namely, they are jointly learned with the encoder at the encoder preparation stage. Once the hyperprior part is trained, we fix this part in the subsequent training procedure. Since the hyperprior part is learned by training the baseline method \cite{balle2018variational} at the base bitrate, it may not achieve promising performance when its input representation is at other bitrates (e.g., $y^{j}$ at the $j$-th supported bitrate). Therefore, we additionally introduce one BAL and one IBAL for the hyperprior part to compensate this representation mismatch.

Different from the structure introduced in Fig.~\ref{fig:CAM}, in our CBANet-HP, each branch of the CAM consists of four deconvolutional layers, and the kernel size of each deconvolutional layer is 5$\times$5. The other settings are the same as those in Fig.~\ref{fig:CAM}.

\begin{figure}[t]
\centering
\includegraphics[width=\textwidth]{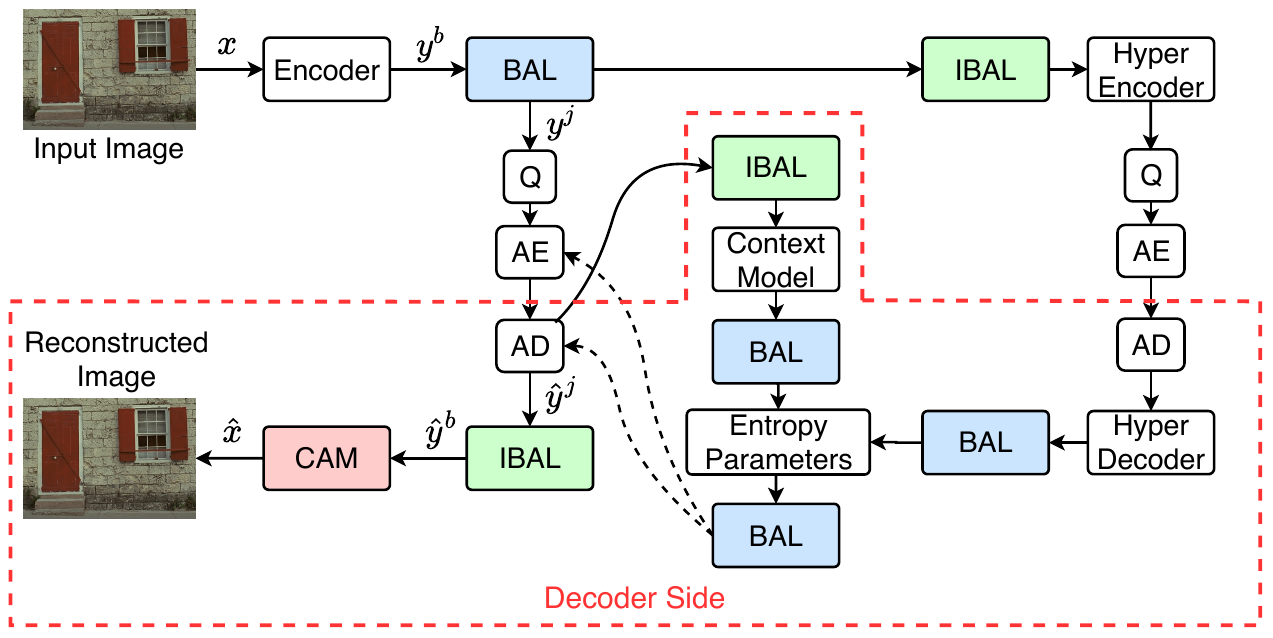}
\caption{Structure of our CBANet-AR, which is built upon the baseline method Minnen et al. (2018)~\cite{minnen2018joint}. Multiple bitrate adaptive layers (BALs) and inverse bitrate adaptive layers (IBALs) are additionally introduced for the hyperprior part (i.e., ``Hyper Encoder'' and ``Hyper Decoder'') and the context part (i.e., ``Context Model'' and ``Entropy Parameters'').}
\label{fig:nips18}
\end{figure}

\textbf{Structure of CBANet-AR: our CBANet based on Minnen et al. (2018).}
For better presentation, we name our CBANet based on the baseline method Minnen et al. (2018)~\cite{minnen2018joint} as \textit{CBANet-AR}. Fig.~\ref{fig:nips18} shows the network structure of our CBANet-AR. Similar to our CBANet-HP, the parameters of the hyperprior part (i.e., ``Hyper Encoder'' and ``Hyper Decoder'' in Fig.~\ref{fig:nips18}) and the context part (i.e., ``Context Model'' and ``Entropy Parameters'' in Fig.~\ref{fig:nips18}) in our CBANet-AR are directly copied from the learned baseline model at the base bitrate. We also fix these two parts for the subsequent training procedure. Since the method in \cite{minnen2018joint} uses a complex structure to generate the entropy model information for the arithmetic encoder and the arithmetic decoder, which includes the context part and the hyperprior part, we additionally introduce two BAL (i.e., one BAL between ``Context Model'' and ``Entropy Parameters'', and one BAL between ``Hyper Decoder'' and ``Entropy Parameters'') in our CBANet-AR to facilitate its training process. The introduction of these two BAL only slightly increases the cost of our CBANet-AR (see Sec.~\ref{Sec:Experiments} for more details), but it can lead to better compression performance.

Similar to our CBANet-HP, each branch of the CAM in our CBANet-AR consists of four deconvolutional layers, and the kernel size of each decovolutional layer is 5$\times$5.

\section{Experiments}
\label{Sec:Experiments}
To demonstrate the effectiveness of our CBANet, we implement our framework based on two popular learning based image compression methods: Ball{\'{e}} et al. (2018)~\cite{balle2018variational} (i.e., CBANet-HP) and Minnen et al. (2018)~\cite{minnen2018joint} (i.e., CBANet-AR). We do not implement our CBANet based on the method \cite{balle2017end} because its performance is not state-of-the-art. We perform the experiments on two benchmark datasets: Kodak~\cite{kodak1993kodak} and Workshop and Challenge on Learned Image Compression (CLIC)~\cite{clic2020dataset}. In this section, we use the number of floating point operations (\#FLOPs) as the criterion for computational complexity measurement, which is commonly used in recent works~\cite{guo2020model,guo2020multi,guo2020channel,zhuang2018discrimination}. Since we focus on the decoder side, we only report the number of FLOPs used in the decoding process when evaluating the computational complexity. 

\textbf{Datasets.}
The Kodak dataset~\cite{kodak1993kodak} consists of 24 uncompressed images with the resolution of $768 \times 512$. 
For the CLIC dataset~\cite{clic2020dataset}, we use the professional dataset in CLIC2020 competition in our experiment, which consists of 41 images.

\subsection{Experiments based on Ball{\'{e}} et al. (2018)}
\label{Sec:IV-A}
\textbf{Implementation details.}
We use 20,745 high-quality images from Flickr.com and take the randomly cropped patches with the resolution of 256$\times$256 as the training data. For the baseline method Ball{\'{e}} et al. (2018)~\cite{balle2018variational}, we follow \cite{balle2018variational} to use 128 channels for the four low bitrates and train the baseline model at the base bitrate by using the Lagrangian multiplier $\lambda$ of 2048. The number of channels $P$ in BAM is set as 192. For the three high bitrates, we follow \cite{balle2018variational} to use 192 channels and train the baseline model at the base bitrate by using the Lagrangian multiplier $\lambda$ of 8192. The number of channels $P$ in BAM is set as 320. We use the Adam optimizer for optimization. In the CAM, we set the total number of branches $K_{max}$ as 3. The number of FLOPs in Branch1, Branch2 and Branch3 take $25\%$, $25\%$ and $50\%$ of the total number of FLOPs in our CAM, respectively. The learning rate and the batch size are set as $1e^{-4}$ and 8 for all the experiments, respectively.

\textbf{Experimental results on Kodak.}
\begin{figure}[t]
\centering
\includegraphics[width=\textwidth]{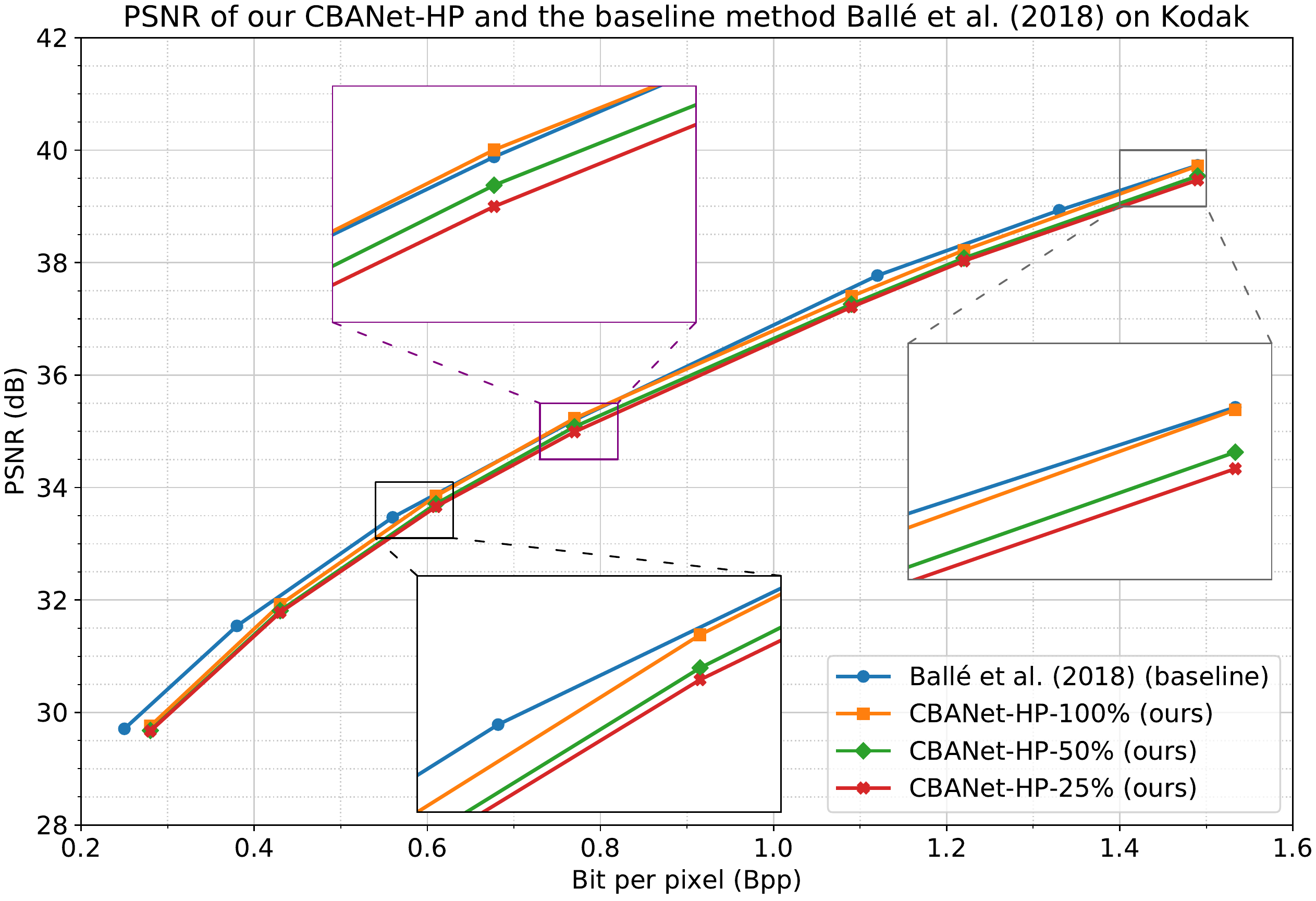}
\caption{PSNR comparison between the baseline method Ball{\'{e}} et al. (2018)~\cite{balle2018variational} and our CBANet-HP at different computational complexity levels on the Kodak dataset.}
\label{fig:iclr18-kodak}
\end{figure}
In Fig.~\ref{fig:iclr18-kodak}, we compare our CBANet-HP with the baseline method Ball{\'{e}} et al. (2018)~\cite{balle2018variational} in terms of peak signal-to-noise ratio (PSNR) on the Kodak dataset for various bitrates and computational complexity levels. In the experimental results, our CBANet-HP when using one, two and three branches in the decoding process are denoted as \textbf{CBANet-HP-25\%}, \textbf{CBANet-HP-50\%} and \textbf{CBANet-HP-100\%}, respectively. Based on these experimental results, we have the following observations:

(1) When using the highest computational complexity level, our CBANet-HP-100\% achieves similar performance when compared with the baseline algorithm \cite{balle2018variational} at most bitrates. It demonstrates the effectiveness of our CBANet-HP for image compression.

(2) Although the average performance of our CBANet-HP-25\% drops about 0.3dB when compared with the baseline method \cite{balle2018variational}, we can significantly reduce the computational complexity by about \textbf{75}\% (e.g., from 61.14G for \cite{balle2018variational} to 16.46G for CBANet-HP-25\% at the low bitrates) to accelerate the decoding process.

(3) For any given bitrate, the PSNR value increases as the computational complexity level increases (i.e., the number of branch increases), which demonstrates the effectiveness of the proposed complexity adaptive module. For example, when bpp is set as 1.49, the PSNR values of our CBANet-HP-25\%, CBANet-HP-50\% and CBANet-HP-100\% are 39.47, 39.54 and 39.72, respectively. 

(4) The performance of our CBANet-HP drops by about 0.5dB at bpp=0.28. We hypothesize that this is mainly because bpp=0.28 is relatively far from bpp=0.77, which is the base bitrate in our implementation. Therefore, it is relatively hard to convert the representation at the base bitrate to that at bpp=0.28. In other words, the BAM is the bottleneck that limits the performance of our CBANet-HP when setting bpp=0.28, which is verified by the results that our CBANet-HP-25\%, CBANet-HP-50\% and CBANet-HP-100\% achieve similar performance when setting bpp=0.28.

To further quantitatively compare the compression performance of our CBANet-HP when using different computational complexity levels, in Table~\ref{tab:bd-iclr18-kodak}, we also provide the BDBR and BD-PSNR values~\cite{bjontegaard2001calculation} when comparing our CBANet-HP with the baseline algorithm \cite{balle2018variational}. Although the BD-PSNR value decreases by 0.13dB for our CBANet-HP-100\% when compared with the method \cite{balle2018variational}, our CBANet-HP can support multiple bitrates and various computational complexity levels by using a single network, which can help us to deploy the learning based image compression systems in different deployment scenarios.
\begin{table}[t]
    \centering
    \small
    \begin{tabular}{c||cc}
    \toprule[1pt]
    Method & BDBR & BD-PSNR \\
    \midrule[1pt]
    \makecell{CBANet-HP-25\%} & 5.82 & -0.31 \\
    \makecell{CBANet-HP-50\%} & 4.92 & -0.27 \\
    \makecell{CBANet-HP-100\%} & 2.56 & -0.13 \\
    \bottomrule[1pt]
    \end{tabular}
    \caption{BDBR(\%) and BD-PSNR(dB) over all bitrates when comparing our CBANet-HP at different computational complexity levels (i.e., with different number of branches) with the baseline method Ball{\'{e}} et al. (2018)~\cite{balle2018variational} on the Kodak dataset.}
    \label{tab:bd-iclr18-kodak}
\end{table}

\textbf{Comparison of storage requirement.}
In Table~\ref{tab:storage-iclr18}, we compare the storage requirement of our CBANet-HP and the baseline method \cite{balle2018variational} when supporting four low bitrates and three computational complexity levels. We have similar observations for the three high bitrates and thus do not provide further analysis. 

\begin{table}[t]
\small
\centering
\caption{The storage sizes for Ball{\'{e}} et al. (2018)~\cite{balle2018variational} and our CBANet-HP when supporting different numbers of low bitrates (i.e., $n=1,2,3,4$) under one or three computational complexity levels.}
\begin{tabular}{c|cc}
\toprule[1pt]
\makecell{\#Supported low bitrates} & 1 & n \\
\midrule[1pt]
\makecell{{Ball{\'{e}} et al. (2018)~\cite{balle2018variational}}\\(for one complexity level)} & 2.53M & (2.53n)M \\
\hline
\makecell{{Ball{\'{e}} et al. (2018)~\cite{balle2018variational}}\\(for three complexity levels)} & 5.93M & (5.93n)M \\
\hline
\makecell{CBANet-HP~(ours)\\(for three complexity levels)} & 2.84M & [2.84+0.17(n-1)]M \\
\bottomrule[1pt]
\end{tabular}
\label{tab:storage-iclr18}
\end{table}
In Table~\ref{tab:storage-iclr18}, the number of parameters of the baseline method \cite{balle2018variational} for supporting one single bitrate and one fixed computational complexity level is 2.53M. One straightforward way to support multiple computational complexity levels is to reduce the number of channels for the baseline method to satisfy different computational complexity constraints and store one deep model for each computational complexity level. When calculating the storage requirement for three computational complexity levels, we need to train three deep models by setting the number of channels in the baseline method \cite{balle2018variational} as 60, 88 and 128, such that the number of FLOPs of the corresponding models are on par with those from our CBANet-HP-25\%, CBANet-HP-50\% and CBANet-HP-100\%, respectively. Therefore, the total number of parameters is 5.93M at each single bitrate, which includes 1.52M, 1.88M and 2.53M for storing the models with 60, 88 and 128 channels, respectively. 

Furthermore, when variable bitrate coding is required in practical applications, the baseline method~\cite{balle2018variational} has to train different models for different bitrates. Therefore, the total number of parameters increases significantly. Specifically, for the single bitrate and three computational complexity levels setting, the number of parameters is 5.93M, while the corresponding number of parameters becomes 23.72M(=5.93$\times$4) for four bitrates and three computational complexity levels setting. In contrast, we only need to store one single model with three BAM (the number of parameters of each BAM is only 0.17M) to support various bitrates under different computational complexity constraints. The number of parameters of our CBANet-HP is 3.35M[=2.84M+0.17M$\times$(4-1)], which achieves a \textbf{85.9}\% reduction in terms of the number of parameters.

\textbf{Practical speedup.} In order to demonstrate the practical speedup after using our proposed method, we report the practical running time of our CBANet-HP and the baseline method \cite{balle2018variational}. Using the machine with one Intel Core i5-7500 CPU, the average latency of the baseline method \cite{balle2018variational}, CBANet-HP-25\%, CBANet-HP-50\% and CBANet-HP-100\% are 633.26ms, 204.59ms, 391.64ms and 629.94ms for decoding one single image on Kodak, respectively. From the results, our CBANet-HP-25\% achieves \textbf{67.69}\% speedup when compared with the baseline method \cite{balle2018variational}. It is worth mentioning that the latency reduction is not linearly correlated to that of the number of FLOPs because the latency is affected by many factors (e.g., the type of CPUs), which is consistent with the observations in many existing works~\cite{he2017channel,wu2019fbnet}.

\begin{figure}[t]
\centering
\includegraphics[width=\textwidth]{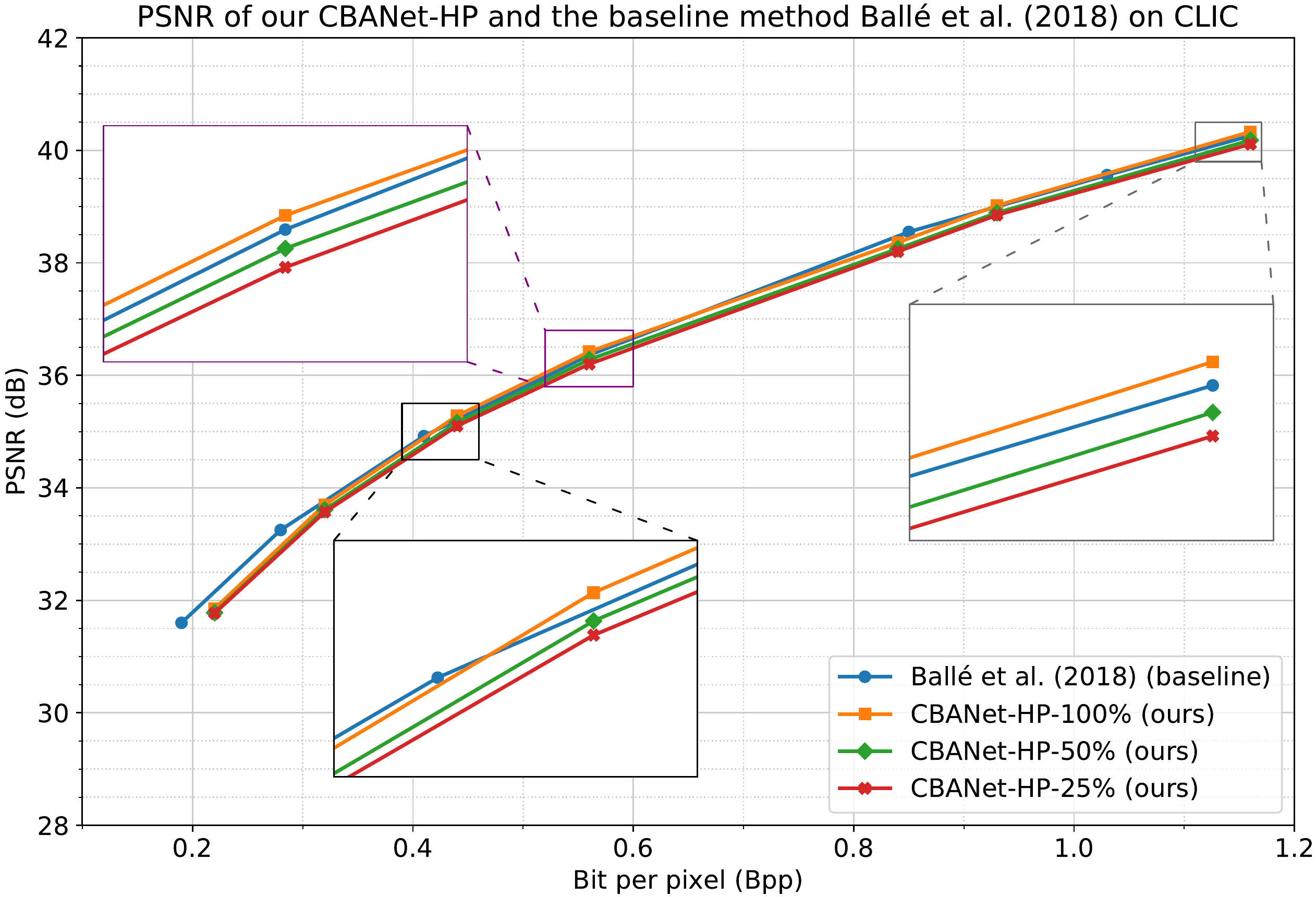}
\caption{PSNR comparison between the baseline method Ball{\'{e}} et al. (2018)~\cite{balle2018variational} and our CBANet-HP at different computational complexity levels on the CLIC dataset.}
\label{fig:iclr18-clic}
\end{figure}

\textbf{Experimental results on CLIC.}
In Fig.~\ref{fig:iclr18-clic}, we compare our CBANet-HP with the baseline method \cite{balle2018variational} in terms of PSNR on the CLIC dataset and we have similar observations as the experiments on Kodak. We would like to highlight that our CBANet-HP-100\% even outperforms the baseline method \cite{balle2018variational} when setting bpp as 0.44, 0.56 and 1.16, which further demonstrates the effectiveness of our CBANet-HP for image compression.

In Table~\ref{tab:bd-iclr18-clic}, we provide the BDBR and BD-PSNR values~\cite{bjontegaard2001calculation} when comparing our CBANet-HP with the baseline algorithm \cite{balle2018variational} on the CLIC dataset. From Table~\ref{tab:bd-iclr18-clic}, the BD-PSNR value only decreases by 0.05dB for our CBANet-HP-100\% when compared with the baseline method \cite{balle2018variational}, which also demonstrates that it is useful to compress the images by using our CBANet-HP.
\begin{table}[t]
    \centering
    \small
    \begin{tabular}{c||cc}
    \toprule[1pt]
    Method & BDBR & BD-PSNR \\
    \midrule[1pt]
    \makecell{CBANet-HP-25\%} & 4.71 & -0.22 \\
    \makecell{CBANet-HP-50\%} & 3.60 & -0.17 \\
    \makecell{CBANet-HP-100\%} & 1.20 & -0.05 \\
    \bottomrule[1pt]
    \end{tabular}
    \caption{BDBR(\%) and BD-PSNR(dB) over all bitrates when comparing our CBANet-HP at different computational complexity levels (i.e., with different number of branches) with the baseline method Ball{\'{e}} et al. (2018)~\cite{balle2018variational} on the CLIC dataset.}
    \label{tab:bd-iclr18-clic}
\end{table}

\subsection{Experiments based on Minnen et al. (2018)}
\textbf{Implementation details.}
Similar to the experiments for CBANet-HP, we train the model at the base bitrate by using the Lagrangian multiplier $\lambda$ of 8192 for the baseline method Minnen et al. (2018)~\cite{minnen2018joint}. We follow \cite{minnen2018joint} to use 320 channels in the model at the base bitrates. The number of channels $P$ in BAM is set as 320. The other settings are the same as those for CBANet-HP except that we use the learning rate of $1e^{-5}$ when training the BAM.

\begin{figure}[t]
\centering
\includegraphics[width=\textwidth]{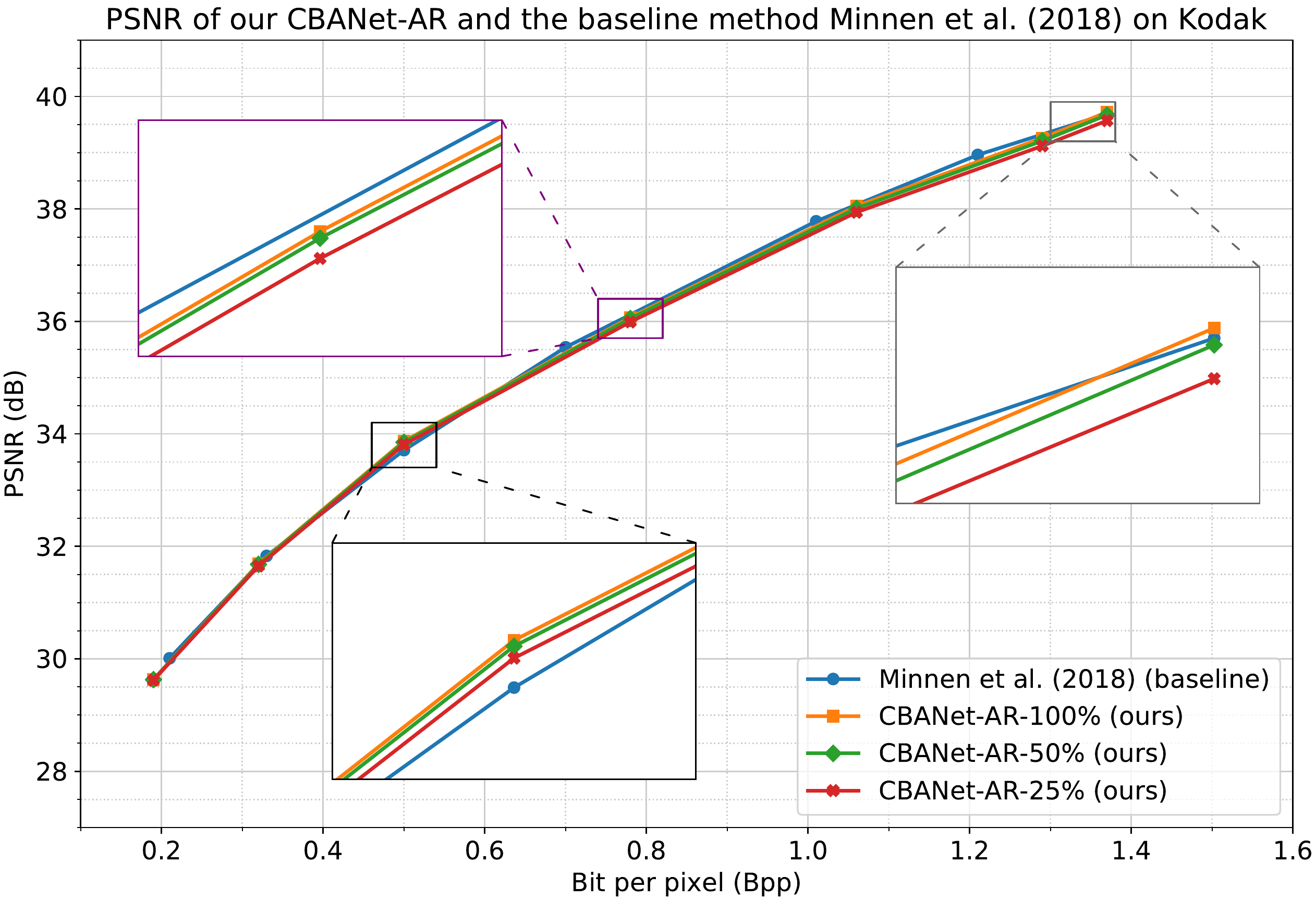}
\caption{PSNR comparison between the baseline method Minnen et al. (2018)~\cite{minnen2018joint} and our CBANet-AR at different computational complexity levels on the Kodak dataset.}
\label{fig:nips18-kodak}
\end{figure}

\textbf{Experimental results on Kodak.}
In Fig.~\ref{fig:nips18-kodak}, we compare our CBANet-AR with the baseline method \cite{minnen2018joint} in terms of PSNR on the Kodak dataset at various bitrates and computational complexity levels. Our CBANet-AR when using one, two and three branches in the decoding process are denoted as \textbf{CBANet-AR-25\%}, \textbf{CBANet-AR-50\%} and \textbf{CBANet-AR-100\%}, respectively. From Fig.~\ref{fig:nips18-kodak}, the performance of our CBANet-AR-100\% is on par with the baseline method Minnen et al. (2018)~\cite{minnen2018joint}. In addition, the performance of our CBANet-AR-25\% only slightly drops when compared with the baseline method \cite{minnen2018joint}. It is worth mentioning that unlike the CBANet-HP method, our CBANet-AR can achieve comparable results at the lowest bitrate (i.e., bpp=0.19) when compared with the baseline method \cite{minnen2018joint}. The results demonstrate the effectiveness of our CBANet-AR for image compression.

In Table~\ref{tab:bd-nips18-kodak}, we report the BDBR and BD-PSNR values~\cite{bjontegaard2001calculation} when comparing our CBANet-AR with the baseline algorithm \cite{minnen2018joint} on the Kodak dataset. From Table~\ref{tab:bd-nips18-kodak}, the BD-PSNR value even increases by 0.01dB for our CBANet-AR-100\% when compared with the baseline method Minnen et al. (2018)~\cite{minnen2018joint}, which shows it is beneficial to compress the images by using our CBANet-AR.
\begin{table}[t]
    \centering
    \small
    \begin{tabular}{c||cc}
    \toprule[1pt]
    Method & BDBR & BD-PSNR \\
    \midrule[1pt]
    \makecell{CBANet-AR-25\%} & 1.15 & -0.06 \\
    \makecell{CBANet-AR-50\%} & 0.24 & -0.01 \\
    \makecell{CBANet-AR-100\%} & -0.16 & 0.01 \\
    \bottomrule[1pt]
    \end{tabular}
    \caption{BDBR(\%) and BD-PSNR(dB) over all bitrates when comparing our CBANet-AR at different computational complexity levels (i.e., with different number of branches) with the baseline method Minnen et al. (2018)~\cite{minnen2018joint} on the Kodak dataset.}
    \label{tab:bd-nips18-kodak}
\end{table}

\textbf{Comparison of storage requirement.}
In Table~\ref{tab:storage-nips18}, we compare the storage requirement of our CBANet-AR and the baseline method \cite{minnen2018joint} when supporting multiple bitrates and various computational complexity levels. We use the same strategy as introduced in Sec. \ref{Sec:IV-A} when calculating the storage requirement for the baseline method \cite{minnen2018joint}. Specifically, when calculating the storage requirement of the method \cite{minnen2018joint} for supporting three complexity levels, we set the number of channels in the baseline method \cite{minnen2018joint} as 136, 215 and 320, such that the number of FLOPs of the corresponding models are on par with those from our CBANet-AR-25\%, CBANet-AR-50\% and CBANet-AR-100\%, respectively. Therefore, the total number of parameters is 65.28M at each single bitrate, which includes 19.08M, 21.19M and 25.01M for storing the models with 136, 215 and 320 channels, respectively. From Table~\ref{tab:storage-nips18}, we observe that our CBANet-AR can significantly reduce the storage requirement when compared with the baseline method \cite{minnen2018joint} for supporting multiple bitrates and various computational complexity levels. For example, the baseline method \cite{minnen2018joint} requires 456.96M(=65.28$\times$7) parameters to support seven bitrates and three computational complexity levels. In contrast, it only takes 36.73M[=26.35+1.73$\times$(7-1)] for our CBANet-AR under this setting, which achieves a \textbf{92.0}\% storage reduction when compared with the baseline method \cite{minnen2018joint}.
\begin{table}[t]
\small
\centering
\caption{The storage sizes for Minnen et al. (2018)~\cite{minnen2018joint} and our CBANet-AR when supporting different numbers of bitrates under one or three computational complexity levels.}
\begin{tabular}{c|cc}
\toprule[1pt]
\makecell{\#Supported bitrates} & 1 & n \\
\midrule[1pt]
\makecell{Minnen et al. (2018)~\cite{minnen2018joint}\\(for one complexity level)} & 25.01M & (25.01n)M \\
\hline
\makecell{Minnen et al. (2018)~\cite{minnen2018joint}\\(for three complexity levels)} & 65.28M & (65.28n)M \\
\hline
\makecell{CBANet-AR~(ours)\\(for three complexity levels)} & 26.35M & [26.35+1.73(n-1)]M \\
\bottomrule[1pt]
\end{tabular}
\label{tab:storage-nips18}
\end{table}

\begin{figure}[t]
\centering
\includegraphics[width=\textwidth]{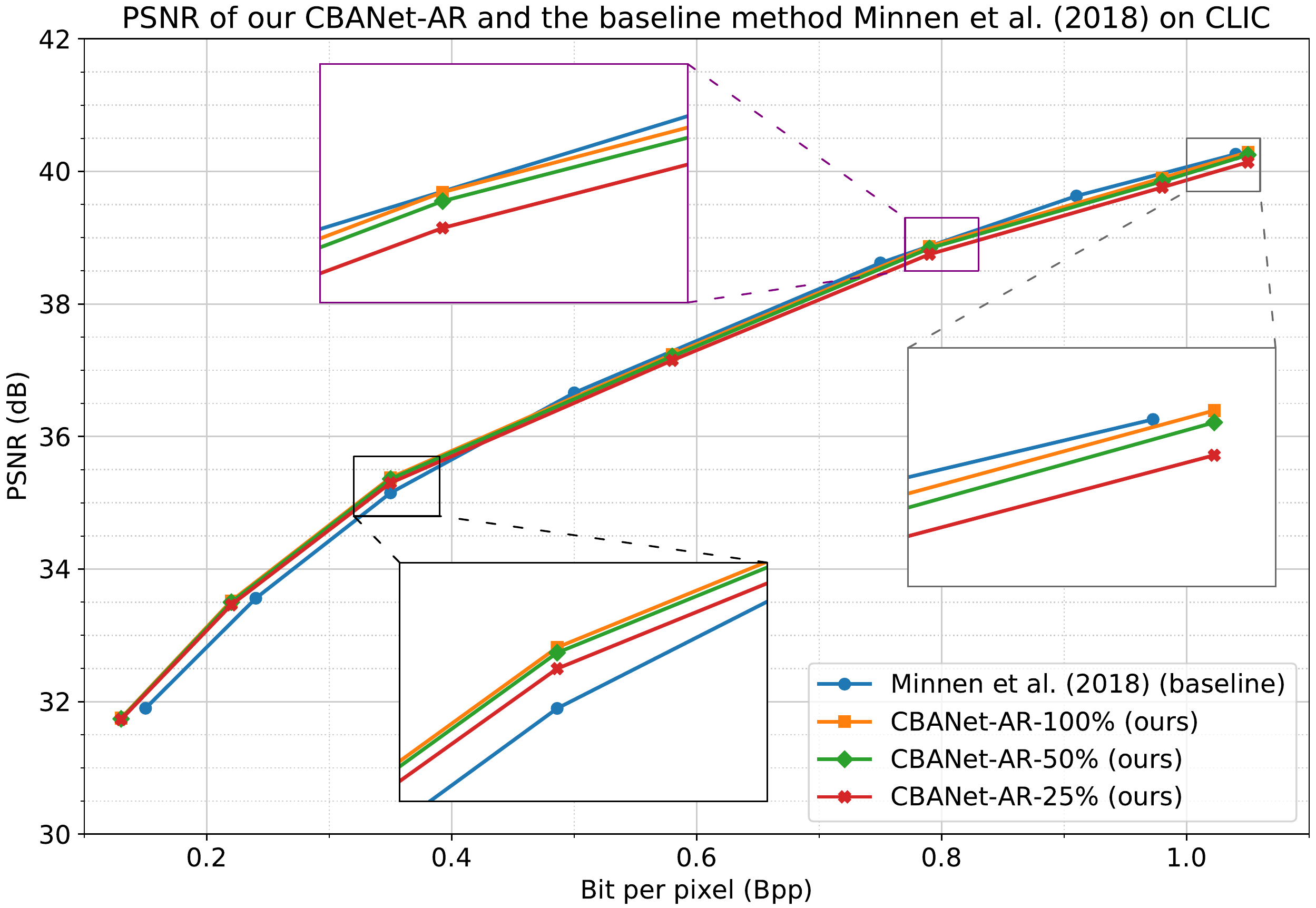}
\caption{PSNR comparison between the baseline method Minnen et al. (2018)~\cite{minnen2018joint} and our CBANet-AR at different computational complexity levels on the CLIC dataset.}
\label{fig:nips18-clic}
\end{figure}

\textbf{Experimental results on CLIC.}
Fig.~\ref{fig:nips18-clic} compares the PSNR of our CBANet-AR and the baseline method Minnen et al. (2018)~\cite{minnen2018joint} on the CLIC dataset. We would like to highlight that our CBANet-AR-25\%, CBANet-AR-50\% and CBANet-AR-100\% even performs better than the baseline method \cite{minnen2018joint} when setting bpp=0.13, bpp=0.22 and bpp=0.35, which demonstrates the effectiveness of our CBANet-AR.

Similar to the experiments on the Kodak dataset, in Table~\ref{tab:bd-nips18-clic}, we report the BDBR and BD-PSNR values~\cite{bjontegaard2001calculation} when comparing our CBANet-AR with the baseline algorithm \cite{minnen2018joint} on the CLIC dataset. From Table~\ref{tab:bd-nips18-clic}, our CBANet-AR-25\%, CBANet-AR-50\% and CBANet-AR-100\% methods outperform the baseline method \cite{minnen2018joint}. Specifically, our CBANet-AR-25\%, CBANet-AR-50\% and CBANet-AR-100\% respectively achieve 0.04dB, 0.10dB and 0.13dB improvement when compared with the baseline method \cite{minnen2018joint}, which indicates that it is useful to compress the images by using our CBANet-AR.
\begin{table}[t]
    \centering
    \small
    \begin{tabular}{c||cc}
    \toprule[1pt]
    Method & BDBR & BD-PSNR \\
    \midrule[1pt]
    \makecell{CBANet-AR-25\%} & -1.07 & 0.04 \\
    \makecell{CBANet-AR-50\%} & -2.39 & 0.10 \\
    \makecell{CBANet-AR-100\%} & -2.97 & 0.13 \\
    \bottomrule[1pt]
    \end{tabular}
    \caption{BDBR(\%) and BD-PSNR(dB) over all bitrates when comparing our CBANet-AR at different computational complexity levels (i.e., with different number of branches) with the baseline method Minnen et al. (2018)~\cite{minnen2018joint} on the CLIC dataset.}
    \label{tab:bd-nips18-clic}
\end{table}

\subsection{Ablation Study and Algorithm Analysis}
\textbf{Effectiveness of our BAM.}
\begin{figure}[t]
\centering
\includegraphics[width=\textwidth]{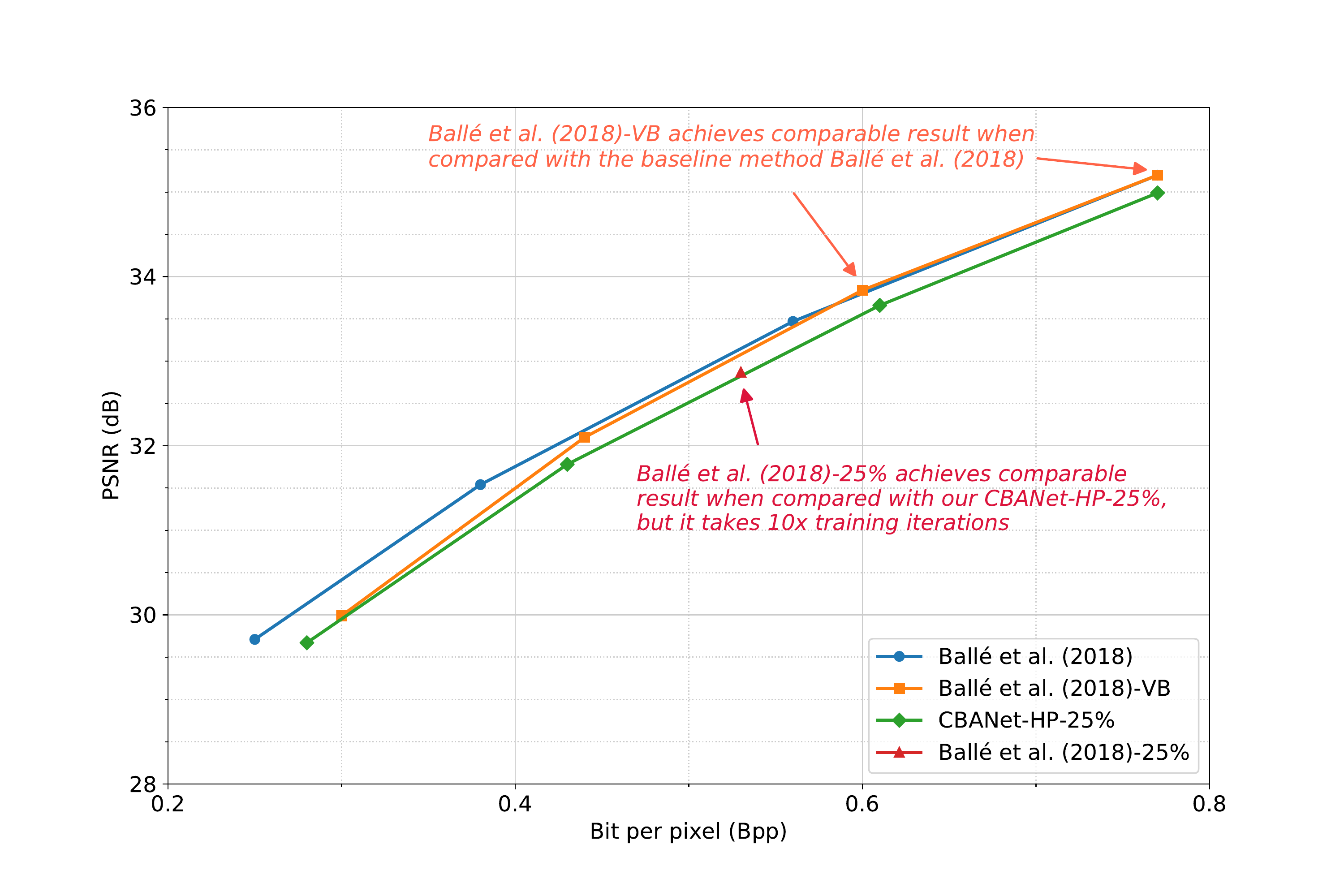}
\caption{PSNR comparison on the Kodak dataset. We report the results of the baseline method Ball{\'{e}} et al. (2018)~\cite{balle2018variational}, our CBANet-HP-25\%, Ball{\'{e}} et al. (2018)-VB, in which we achieve variable bitrate image compression by only using our BAM without introducing our CAM, and Ball{\'{e}} et al. (2018)-25\%, in which we directly train the codec in \cite{balle2018variational} from scratch by using a low complexity decoder.}
\label{fig:ablation}
\end{figure}
To demonstrate the effectiveness of our BAM for transferring the representation from the base bitrate to those at different bitrates, we take the method Ball{\'{e}} et al. (2018)~\cite{balle2018variational} as an example and perform the experiments on the Kodak dataset to only transfer the representation without introducing the CAM in our CBANet-HP to support multiple bitrates. In this case, we firstly train an encoder-decoder pair at the base bitrate by using the method \cite{balle2018variational}. Then we introduce our BAM in this codec for variable bitrate image compression without replacing the decoder with the CAM. The result is referred to as \textit{Ball{\'{e}} et al. (2018)-VB} in Fig.~\ref{fig:ablation}. From Fig.~\ref{fig:ablation}, we observe that the Ball{\'{e}} et al. (2018)-VB method is comparable with the baseline method \cite{balle2018variational} at most bitrates, which demonstrates that it is effective to use our BAM to support variable bitrate image compression. Moreover, the performance of the Ball{\'{e}} et al. (2018)-VB method drops about 0.4dB at bpp=0.30 when compared with the baseline method \cite{balle2018variational}, which again demonstrates that the BAM limits the performance of our CBANet-HP at the lowest bitrate.



We also report the number of FLOPs of the BAM in our CBANet-HP. Taking our CBANet-HP-100\% at the low bitrates as an example, the number of FLOPs of the BAM and our CBANet-HP-100\% are 0.26G and 61.52G, respectively. The number of FLOPs of our BAM only takes \textbf{0.42}\% of the total number of FLOPs, which indicates that our CBANet-HP can support more bitrates by slightly increasing the storage cost.

\textbf{Comparison with the alternative method by training from scratch.}
One may ask the question: How is the performance if we directly train the codec with a low complexity decoder (i.e., with less number of FLOPs) by using the existing methods? To address this concern, we take the method Ball{\'{e}} et al. (2018)~\cite{balle2018variational} as an example and perform the experiment to use less number of FLOPs in the decoder by setting the number of channels in the decoder as 58. We set the encoder structure the same as that in the baseline method \cite{balle2018variational}. In this case, the number of FLOPs for the decoder is on par with our CBANet-HP-25\%. We directly train the codec with the low complexity decoder from scratch and the result is referred to as \textit{Ball{\'{e}} et al. (2018)-25\%} in Fig.~\ref{fig:ablation}. From the results, we observe that the performance of the Ball{\'{e}} et al. (2018)-25\% method is comparable with our CBANet-HP-25\% approach. However, we would like to highlight that it takes 3,000,000 iterations for the Ball{\'{e}} et al. (2018)-25\% method to learn one model at each bitrate. On the other hand, it only takes 300,000 iterations in our CBANet-HP-25\% to learn one BAM for supporting one additional bitrate after we train the model at the base bitrate, which is \textbf{10}$\times$ less than that in the Ball{\'{e}} et al. (2018)-25\% method.


\section{Conclusion}
In this work, we have introduced a rate-distortion-complexity optimization based framework called Complexity and Bitrate Adaptive Network (CBANet) to achieve complexity and bitrate adaptive deep image compression by using one single network. To address this challenging new research problem, our CBANet seamlessly integrates a newly proposed complexity adaptive module and a new bitrate adaptive module. In contrast to the existing learning based deep image compression approaches that need to train different models for different bitrates and different computational complexity constraints, our CBANet can support multiple bitrate image compression under various computational complexity constraints by using one single network. Extensive experiments on two benchmark datasets demonstrate the effectiveness of our proposed CBANet.


%





\ifCLASSOPTIONcaptionsoff
  \newpage
\fi



%
{\small
\bibliographystyle{IEEEtran}
\bibliography{egbib}
}

%

\newpage
\begin{IEEEbiography}[{\includegraphics[width=1in,height=1.25in,clip,keepaspectratio]{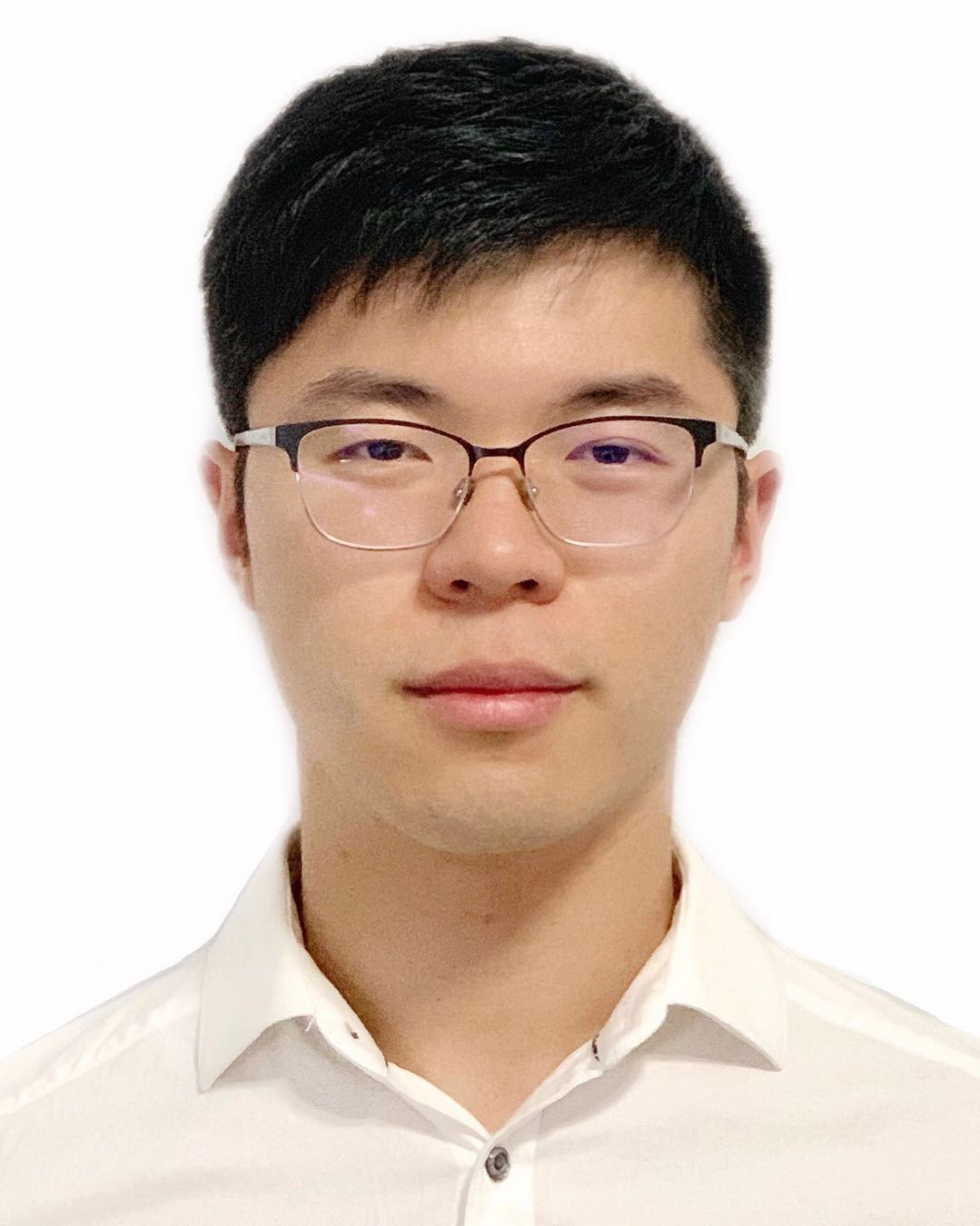}}]{Jinyang Guo}
received the BE degree in School of Electrical Engineering and Telecommunications from the University of New South Wales in 2017. He is currently pursuing the PhD degree in the School of Electrical and Information Engineering, the University of Sydney. His research interests include deep model compression and its applications on computer vision.
\end{IEEEbiography}
\vspace{-15mm}
\begin{IEEEbiography}[{\includegraphics[width=1in,height=1.25in,clip,keepaspectratio]{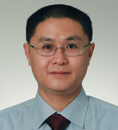}}]{Dong Xu}
 received the BE and PhD degrees from University of Science and Technology of China, in 2001 and 2005, respectively. While pursuing the PhD degree, he was an intern with Microsoft Research Asia, Beijing, China, and a research assistant with the Chinese University of Hong Kong, Shatin, Hong Kong, for more than two years. He was a post-doctoral research scientist with Columbia University, New York, NY, for one year. He worked as a faculty member with Nanyang Technological University, Singapore. Currently, he is a professor and chair in Computer Engineering with the School of Electrical and Information Engineering, the University of Sydney, Australia. His current research interests include computer vision, statistical learning, and multimedia content analysis. He was the co-author of a paper that won the Best Student Paper award in the IEEE Conferenceon Computer Vision and Pattern Recognition (CVPR) in 2010, and a paper that won the Prize Paper award in IEEE Transactions on Multimedia (T-MM) in 2014. He is a fellow of the IEEE.
\end{IEEEbiography}
\vspace{-15mm}
\begin{IEEEbiography}[{\includegraphics[width=1in,height=1.25in,clip,keepaspectratio]{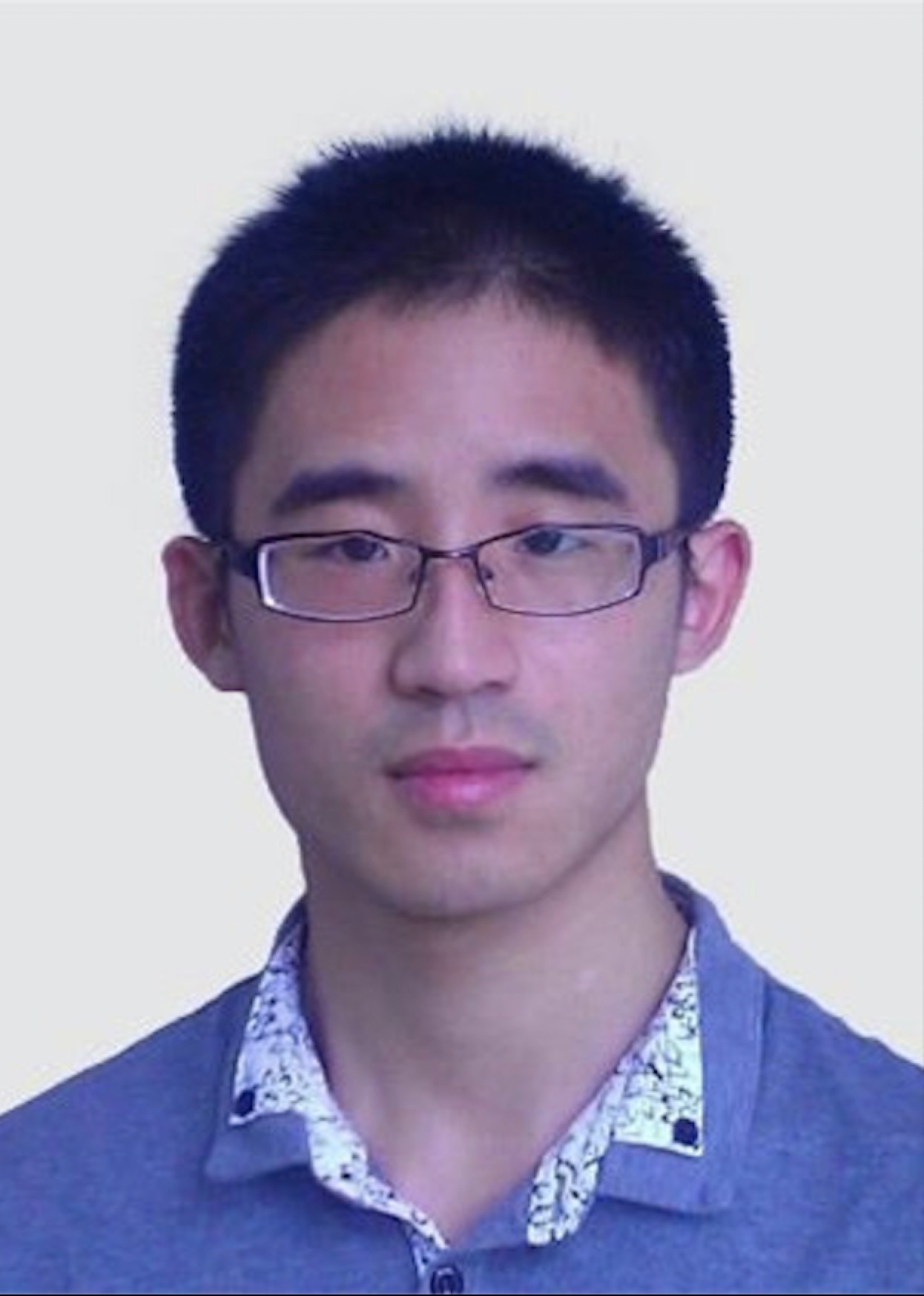}}]{Guo Lu} received his PhD degree from Shanghai Jiao Tong University in 2020 and the B.S. degree from Ocean University of China in 2014. Currently, he is an assistant professor with the School of Computer Science,  Beijing Institute of Technology, China. His research interests include image and video processing, video compression and computer vision.  His works have been published in top-tier journals and conferences (e.g., T-PAMI, T-IP, CVPR and ECCV).
\end{IEEEbiography}







\end{document}